\documentclass
[12pt]
{iopart}
\usepackage{epsfig}

\begin{document}
\title{Energies of sp$^2$ carbon shapes with pentagonal 
disclinations and elasticity theory}
\author{Antonio \v{S}iber}
\address{Institute of Physics, P.O. Box 304, 10001 Zagreb, Croatia}

\ead{asiber@ifs.hr}

\begin{abstract}
Energies of a certain class of fullerene molecules (elongated, contracted, and 
regular icosahedral fullerenes) are numerically calculated using a microscopic 
description of carbon-carbon bonding. It is shown how these results can be 
interpreted and comprehended using the theory of elasticity that describes 
bending of a graphene plane. Detailed studies of a 
wide variety of structures constructed by application of the same general principle 
are performed, and analytical expressions for energies of such structures are 
derived. Comparison of numerical results with the predictions of 
a simple implementation of elasticity theory confirms the usefulness of 
the latter approach.
\end{abstract}
\maketitle

\begin{center}
Published in Nanotechnology {\bf 17}, 3598 (2006)
\end{center}
\section{Introduction}

\label{sec:intro0}

Recent years have been marked by an accented interest of scientists in molecules and 
materials made exclusively of carbon atoms (C). This interest has been 
initiated by discoveries of fullerenes \cite{discofull} and 
carbon nanotubes \cite{discoIjima}, and is fueled by many different nanoscopic 
shapes and structures made of carbon atoms that are continually appearing either as 
a result of experimental studies \cite{newshapesexp,newshape2,newshape3} 
or imaginative theoretical 
constructions \cite{imaginativetheor,imaginative2,imaginative3,negcurvature}. 
Structures of interest to this 
work are those in which carbon atoms are sp$^2$ hybridized, as in the 
graphene layers in graphite. Fullerenes and carbon nanotubes 
fit in this category, but there are many other different shapes and objects that 
can be constructed. The multitude of possible shapes is a result of pronounced 
anisotropy of carbon-carbon interactions and the relatively low energy needed 
to transform a hexagonal carbon ring into a pentagonal one. Inclusion of 
heptagonal rings allows for creation of additional, 
negative-Gaussian-curvature shapes \cite{negcurvature} that 
are not a subject of this article. Carbon shapes are of interest since 
the same elementary constituents (sp$^2$ hybridized carbon atoms) can be 
''assembled'' in a variety of stable shapes that will not spontaneously 
decay or transform in some other shapes once created. The resulting shapes 
and structures may differ in 
their functionality, whatever their purpose may be. The assembly of 
simple elementary pieces into a possibly ''engineered'' or ''designed'' 
shape is a dream of nanotechnologists, and the carbon structures are thus 
the ideal benchmark system. Nanoscale shapes of biological interest are 
often self-assembled - a prime example is the selfassembly of viral 
coatings from individual proteins that make it \cite{Frenkel}. Even more 
intriguingly, the symmetries and shapes of icosahedral viruses are directly related 
to symmetries of ''gigantic'' fullerenes \cite{Kroto,Lidmar,Bruinelast}. 
Both types of structures 
are characterized by pentagonal ''disclinations'' \cite{Seung} 
(pentagonal carbon rings in fullerenes \cite{Kroto} vs. pentameric protein 
aggregates in viruses \cite{Virrev}) in crystalline 
sheet, and both can be constructed from the triangulation of an icosahedron. 
This may be a coincidence, but may also point to more general and fundamental 
principles that may still be discovered in the study of carbon structures 
and shapes. Viruses, as carbon shapes, also show the property of multiformity. 
The same viral coating proteins may assemble in icosahedral shells of varying 
symmetry, but may also make spherocylindrical, conical particles or 
even protein sheets, depending on the conditions \cite{Bruinelast,Virrev}. 
Similar features are found in sp$^2$ carbon shapes. Although various shapes 
can be, more or less easily imagined, 
this does not mean that they should have large binding energies, or that they 
should be bound at all. Therefore, it seems pragmatic to have a 
rule of thumb estimate for their energy, based solely on their imagined shape. 
Theory of elasticity provides an excellent framework for such estimates and 
has been in the past successfully applied for this purpose 
\cite{Lidmar,Bruinelast,tersselast,brennelast}. The aim of this article is to show how 
the energies of the various shapes that can be imagined may be estimated 
from simple relationships that are quite accurate when applied 
properly. In certain aspects, this article is a continuation of efforts 
started in Refs. \cite{tersselast} and \cite{brennelast}, but 
differs from them in that it studies a certain class of convex carbon shapes that 
include both capped carbon nanotubes and fullerenes as special limiting cases. It 
also applies the theory of elasticity in more difficult circumstances in which 
the symmetry of shapes is reduced. 

In Sec. \ref{sec:sec1} I shall present the geometric construction of 
a certain class of shapes 
that is of interest to this article. In Sec. \ref{sec:sec2} I shall describe the 
procedure that is used to find minimal energy shapes based on the geometric 
construction discussed in Sec. \ref{sec:sec1}. The procedure is based on the implementation 
of conjugate gradient technique \cite{CG} in combination with the latest Brenner's 
potential for the description of carbon-carbon bonding \cite{Brennerpot}, which 
provides an excellent opportunity to further examine the predictions of 
this relatively recently proposed potential, although the main message 
of this article is totally independent on the form of the potential used. 
Section \ref{sec:sec3} contains numerical results of this article, 
together with elements of elasticity theory that are applied in 
the analysis of these results. It will be demonstrated how a 
simple characterization of the graphene elasticity together with 
the knowledge of the energy for creation of pentagonal disclination enables 
a reliable estimate of energies of various convex shapes. Section \ref{sec:sec4} 
summarizes and concludes the article.

\section{Construction of shapes of interest}
\label{sec:sec1}
\begin{figure}[ht]
\centerline{
\epsfig {file=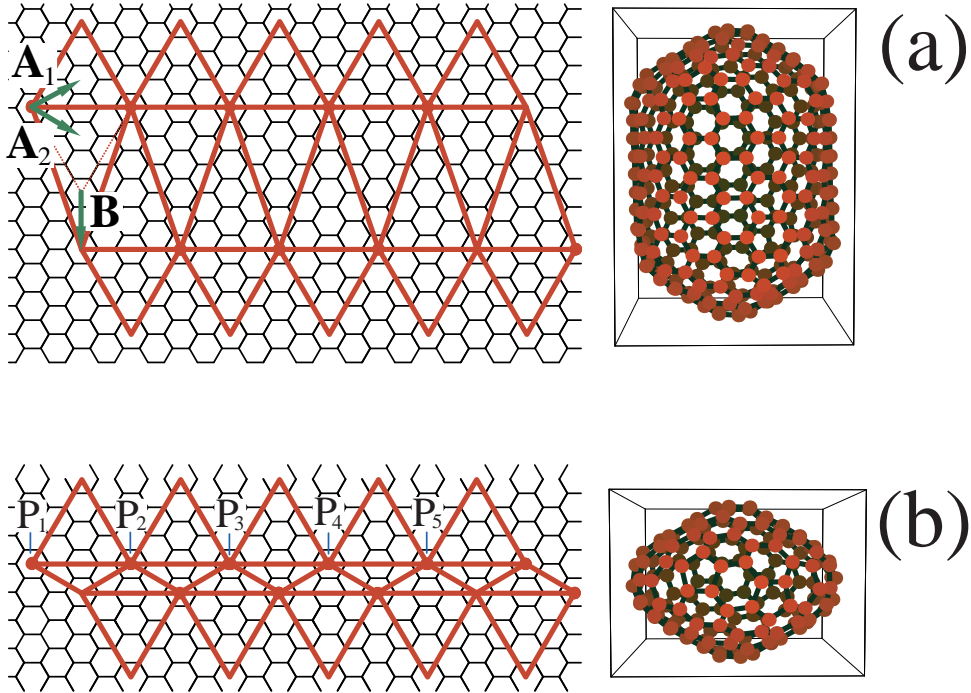,width=12.9cm}
}
\caption{Geometric construction of the ''armchair'' shapes of interest. 
The shape is cut out from the graphene plane and folded 
in an elongated (shape in panel (a)) or contracted (shape in panel (b)) or 
regular (not shown) icosahedron. Note that the points P$_1$, ..., P$_5$ 
(panel (b)) are situated in polygon vertices of a (large) pentagon 
once the shape is folded.
}
\label{fig:fig1}
\end{figure}

The shapes of interest to this work can be constructed as depicted in 
Fig. \ref{fig:fig1}.
The procedure is to cut out the outlined shape from 
the graphene plane and fold it into a polyhedron. The shapes that 
shall be considered can be loosely termed as elongated, regular or contracted 
icosahedra. Note that the construction results in twelve pentagonal 
carbon rings situated in vertices of (contracted or elongated) icosahedron. 
The equilibrated, minimal energy shapes obtained from the pieces of 
cut-out graphene plane are also displayed in 
Fig. \ref{fig:fig1}. The numerical procedure that was used to obtain these 
shapes will be discussed in detail in Sec. \ref{sec:sec2}. At this point, 
not only how the elongated shapes resemble carbon nanotubes, while the highly 
contracted shapes look almost like two cones glued together at their bases 
- there are two pentagonal carbon rings at the ''poles'' of the shape 
(upper and lower points in shape depicted in panel (b) of Fig. \ref{fig:fig1}), 
and the remaining ten pentagonal carbon rings are arranged around an 
''equator'' of the shape, in the polygon vertices of two parallel 
large pentagons rotated 
with respect to each other by an angle of $\pi / 5$ (the vertices of one 
of those pentagons are denoted by points P$_1$,...,P$_5$ in Fig. \ref{fig:fig1}).
The shapes can be uniquely characterized by the lengths of two 
vectors, ${\bf A}_1$ and ${\bf B}$, since $|{\bf A}_1| = |{\bf A}_2|$ 
(see Fig. \ref{fig:fig1}). The lengths are integer multiples of the 
distance between the centers of neighbouring carbon hexagonal rings, i.e. 
\begin{eqnarray}
|{\bf A}_1| &=& |{\bf A}_2| = m d, \nonumber \\
|{\bf B}| &=& n d, \nonumber \\
d &=& a \sqrt{3}
\end{eqnarray}
where $a$ is the nearest neighbor C-C distance in graphite. I shall adopt the 
convention that negative values of $n$ correspond to contracted shapes, so 
that the shapes in Fig. \ref{fig:fig1} can be described by 
$(m=2, n=2)$ (elongated icosahedron) and $(m=2, n=-2)$ (contracted 
icosahedron). The (regular) icosahedral fullerenes are obtained for $n=0$. The 
condition that has to be fulfilled by $m$ and $n$ integers is $n \ge -m$. For 
$n \gg m$, the construction results in capped armchair $(5m,5m)$ single-wall 
carbon nanotubes and that is the reason for giving the 
attribute ''armchair'' to the chosen subset of shapes. The total number of 
carbon atoms ($N$) in the constructed shapes is
\begin{equation}
N = 60m^2 + 20mn.
\label{eq:totnum}
\end{equation}
Similar geometrical construction can be used to generate to ''zig-zag'' 
or chiral (helical) shapes \cite{Terrones}. However, as 
the main intention of this work is in application of a continuum elasticity 
theory, the precise structure and atomic symmetry of the shape is of no 
importance, and only its shape matters in this respect. It should be 
noted here that the construction results in {\em single-shell} or 
{\em single-wall} carbon shapes/molecules. All the effects 
that are specific to multishell structures, in particular 
van der Waals interaction between the different shells are not 
treated in this work. For the application of the continuum 
elasticity theory in presence of van der Waals interactions 
and/or surrounding elastic medium see e.g. Refs. \cite{referee1,referee2}.

\section{Finding the minimum-energy shapes}
\label{sec:sec2}

The mathematically constructed folded polyhedra will not 
represent the minimum energy shapes. Carbon atoms in a minimum-energy 
structure should be relaxed 
so that the nearest-neighbour C-C distances are close to their 
optimum value energywise. To account for these effects, one should have a reliable 
description of the energetics of carbon bonding. In this work, the C-C 
interactions are modeled using a relatively recent 
second-generation reactive empirical bond order potential by 
Brenner et al \cite{Brennerpot}. In this model, the potential energy 
($E_b$) of a carbon structure is given by 
\begin{equation}
E_b = \sum_i \sum_j \left[ V^R (r_{ij}) - b_{ij}V^A(r_{ij}) \right],
\label{eq:modelpotential}
\end{equation}
where $r_{ij}$ is the distance between nearest-neighbour atoms $i$ and 
$j$, $V^R$ and $V^A$ are pair-additive repulsive and attractive 
attractions, respectively, and $b_{ij}$ is the bond order between 
the atoms $i$ and $j$ that approximately accounts for the anisotropy of 
C-C bonding and for the many-body 
character of C-C interactions by including effectively the three-body 
contributions to $E_b$. The detailed description of the potential, together 
with all relevant parameters can be found in Ref. \cite{Brennerpot}. 
Similar potential models have been successfully employed in 
previous research of carbon structures (see e.g. Refs. 
\cite{tersselast,brennelast}). 

The geometrically constructed, folded shape is used as the initial 
guess of the structure. The local minimum of $E_b$ in the configurational 
space spanned by coordinates of all carbon atoms is found using the 
conjugate gradient technique described in Ref. \cite{CG}. The conjugate 
gradient technique optimizes all the coordinates 
at once, proceeding through a sequence of steps along the ''noninterfering'' 
or ''conjugate'' direction in the configurational space, so that a 
minimization along a particular direction does not spoil the effect 
of minimization in other conjugate directions. Similar 
procedures have been used for the same purposes in 
Refs. \cite{Lidmar,Bruinelast,Seung}. 

\section{Energies of shapes and considerations based on elasticity theory}
\label{sec:sec3}
\subsection{Infinitely long carbon cylinders (non-capped carbon nanotubes)}

\begin{figure}[ht]
\centerline{
\epsfig {file=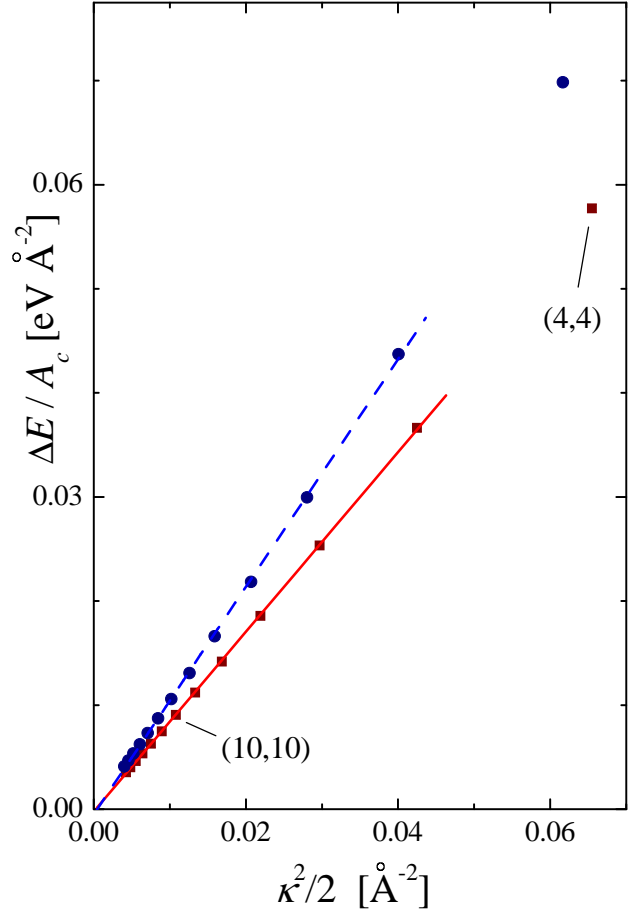,width=9cm}
}
\caption{Excess energy per unit area as a function of squared cylinder 
(armchair carbon nanotube) curvature for 
the Tersoff potential \cite{Terspot} (circles) and the Brenner potential
\cite{Brennerpot} (squares). The results for (4,4) and (10,10) armchair 
carbon nanotubes in this graph are denoted.}
\label{fig:fig2}
\end{figure}

Conceptually the simplest shape that can be constructed from the graphene plane 
is the open-ended (non-capped) carbon nanotube. It does not result from 
the construction described in Sec. \ref{sec:sec1}, but shall be examined 
first to enable a simple insight in the application of elasticity 
theory to the shapes of interest. I shall consider infinitely long single-walled 
armchair carbon nanotubes. The energetics of such shapes can be studied by 
applying the (one-dimensional) periodic boundary conditions to the 
carbon-nanotube unit cell (ring or several rings of atoms), so that the effective number 
of independent coordinates that have to be optimized with respect to 
total energy of the nanotube is relatively small. Figure \ref{fig:fig2} 
displays the energy of infinite, armchair-type carbon cylinders as a function of 
the squared cylinder curvature given by $\kappa = 1/\langle R_{cyl} \rangle$. The 
mean radius of the cylinder, $\langle R_{cyl} \rangle$ was calculated from 
the relaxed atomic coordinates as
\begin{equation}
\langle R_{cyl} \rangle = \frac{\sum_i \sqrt{x_i^2 + y_i^2}}{N},
\label{eq:cylrad}
\end{equation}
where $N$ is the total number of atoms in the unit cell, ${\bf r}_i 
= (x_i, y_i, z_i)$ is the position vector of the $i$-th carbon atom, 
and the cylinder axis coincides with the $z$-axis. The cylinder 
radius calculated from Eq. (\ref{eq:cylrad}) is always very close 
to the analytical prediction for tube radius, $R = a / \{2 \sin[\pi/ (3m)] \}$. 
For example, for $(10,10)$ carbon nanotubes, Eq. (\ref{eq:cylrad}) and 
the numerically calculated values of ${\bf r}_i$ yield nanotube 
radius of 6.802 \AA, while the analytical prediction is 6.794 \AA. The 
agreement becomes slightly worse with the decrease in the nanotube radius.
The energies displayed in 
Fig. \ref{fig:fig2} are in fact excess energies {\em per unit area} of the 
structure (denoted by $A_c$; $A_c = 3 a^2 \sqrt{3} / 4$) and were calculated as
\begin{equation}
\frac{\Delta E}{A_c} = (E_{cylinder}^{per \; atom} - E_{graphene}^{per \; atom}) 
\frac{4}{3 a^2 \sqrt{3}},
\end{equation}
where $E_{cylinder}^{per \; atom}$ and $E_{graphene}^{per \; atom}$ are the 
energies {\em per atom} in the cylinder and in the graphene plane, respectively. 
The calculations were carried out for two versions of the carbon-carbon potential 
which obviously predict different slopes of the 
dependence of the excess energy on squared curvature, but note that both 
curves are {\em linear} up to very large curvatures i.e. small radii of 
carbon nanotubes. Calculations with the early Tersoff potential \cite{Terspot} predict 
that somewhat larger energies are required to bend a piece of graphene plane 
in a cylinder with respect to those obtained by using the latest 
Brenner potential \cite{Brennerpot}. For the energy per unit atom in the 
graphene plane the two potentials predict $E_{graphene}^{per \; atom} = -7.39552$ eV 
(Tersoff's potential) and $E_{graphene}^{per \; atom} = -7.39494$ eV (Brenner's 
potential). More important difference between the two potentials is in the equilibrium C-C 
bond lengths they predict ($a=1.4605$ \AA \hspace{0.8mm} and $a=1.4204$ \AA 
\hspace{0.8mm} for the Tersoff and Brenner potential, respectively). The 
slopes of the lines shown in Fig. \ref{fig:fig2} ($c_0$) were 
obtained from fitting the calculated results pertaining to 
(5,5),(6,6),...,(16,16) carbon nanotubes, and for the Brenner and the 
Tersoff potential they are 0.863 $\pm$ 0.003 eV and 1.093 $\pm$ 0.006 eV, 
respectively (Tersoff obtained $c_0 = 1.02$ eV in Ref. \cite{tersselast}).

The demonstrated linear dependence of the excess energy per unit area on squared 
(mean) curvature can be easily understood from the examination of the geometry of 
bent network of graphene bonds. Consider a piece of graphene plane bent on 
a large cylinder of radius $R \gg a$. The most important change in the 
energetics of carbon network is due to a {\em change in angles between the 
bonds}, while the bond lengths remain the same as in graphene. Only 
the bond order term, $b_{ij}$ depends on the angles between bonds 
(see Eq. (\ref{eq:modelpotential}) and Refs. \cite{Brennerpot,Terspot}). 
By purely geometrical considerations it is easy to derive how the angles 
between the bonds change when the graphene plane is rolled onto a cylinder. 
If one assumes the plane to be inextensional, and considers large 
cylinder radii ($a/R \ll 1$) the change in energy per $i$-th atom is given by
\begin{equation}
\Delta E_i = \frac{1}{2} V^A (a) \left [
\frac{\partial b_{ij}}{\partial \cos (\theta_{ijk})} \right ]_{-1/2} 
\left ( \frac{3a}{4R} \right )^2,
\end{equation}
which can also be written as 
\begin{equation}
\frac{\Delta E}{A_c} = \frac{1}{2} V^A(a) \left[ 
\frac{\partial b_{ij}}{\partial \theta_{ijk}}
\right ]_{2 \pi / 3} \frac{\kappa^2}{2},
\label{eq:atomic1}
\end{equation}
where $j$ is the index of one of the three nearest neighbours of $i$-th atom, 
$k \ne j$ is any of the remaining two neighbors of $i$, 
and the angle between the $i-j$ and $i-k$ bonds is $\theta_{ijk}$ 
\footnote{The bond order function $b_{ij}$ includes 
summation over $k$, the remaining two neighbors of $i$, $k \ne j$ 
(see Refs. \cite{Brennerpot,Terspot}). When evaluating 
the derivative in Eq. (\ref{eq:atomic1}), all the indices ($i,j,k$) are to 
be considered as fixed, but the summation over $k$ in $b_{ij}$ of course remains.}
Explicitly written factor of 1/2 is a consequence of the fact that 
the energy in the Tersoff-Brenner potentials is situated in C-C bonds, 
and that each bond is shared by {\em two} atoms. Equation (\ref{eq:atomic1}) is most 
easily derived by considering armchair or zig-zag carbon nanotubes, since in 
these cases one of the bonds is either parallel (zig-zag) or perpendicular 
(armchair) to the cylinder axis, but for $a/R \ll 1$, the relation holds 
also for helical nanotubes, i.e. for small curvatures the bending energies 
do not depend on the symmetry of the carbon nanotube. Note again that 
the analytical expression (\ref{eq:atomic1}) accounts for the excess energy only in 
the limit of small curvatures $\kappa$ i.e. for radii of curvature that 
are large with respect to the graphene lattice constant.

Equation (\ref{eq:atomic1}) shows that the excess energy per unit area is proportional 
to the squared curvature, in accordance with the results produced by 
calculations displayed in Fig. \ref{fig:fig2} (note, however, that 
only a curvature along one direction is accounted for by Eq. (\ref{eq:atomic1}) 
which is sufficient for cylindrical surface). 
It furthermore predicts that 
the slope of this dependence is given by the properties of the interatomic 
potential (intriguingly, the factor of proportionality does not depend on 
the form of the repulsive part of potential for the class of bond order 
potentials given in Refs. \cite{Brennerpot,Terspot}). The slope $c_0$ can 
be easiliy read out from Eq. (\ref{eq:atomic1}), and is calculated 
to be $c_0 = 0.8301$ eV and $c_0 =1.0152$ eV for Brenner's and Terssof's 
potentials respectively. This is fairly close to the slopes obtained 
numerically. The small difference can be attributed 
to the fact that Eq. (\ref{eq:atomic1}) is valid only for small curvatures, 
(or large radii, $a/R \ll 1$), i.e. it includes only the lowest order 
contribution of curvature (quadratic). For tubes of small radii, higher order 
terms are expected to be of importance. In fact, if one restricts the 
linear fit in Fig. \ref{fig:fig2} to smaller curvatures 
(from (12,12) to (16,16) nanotubes, $a/R < 0.175$) one finds that the 
slopes are 0.839 eV and 1.036 eV, which is in significantly better agreement with 
the prediction of Eq. (\ref{eq:atomic1}). Note that for (5,5) nanotubes, 
$a/R = 0.414$, which means that the tubes of such small radii are already in the 
region where condition $a/R \ll 1$ does not hold, and the predictions 
of Eq. (\ref{eq:atomic1}) are not expected to be very accurate. 

The key conclusion of this subsection is that the excess energy per 
carbon atom ($\Delta E$) of an infinitely long carbon cylinder can be reliably 
estimated from 
\begin{equation}
\frac{\Delta E}{A_c} = c_0 \frac{\kappa^2}{2},
\label{eq:elastica1}
\end{equation}
where $c_0$ is the elastic constant of the graphene plane related to the 
energetics of its bending (the bending rigidity), which, 
for infinitensimal curvatures, can be calculated from the knowledge of 
the interatomic interactions in graphene as in Eq. (\ref{eq:atomic1}). This 
equation holds irrespectively of the details of the interatomic 
interactions, which only change the {\em value} of $c_0$ but not 
the {\em functional dependence} displayed in Eq. (\ref{eq:elastica1}), which 
was illustrated by the examination of energetics predicted by 
two different models of C-C interactions \cite{Brennerpot,Terspot}. 
Equation (\ref{eq:elastica1}) can also be understood without reference to 
atomic structure of the carbon 
nanotube by interpreting its left hand side as the bending energy per unit 
area of the material.

\subsection{Icosahedral fullerenes}
\label{icos}

\begin{figure}[ht]
\centerline{
\epsfig {file=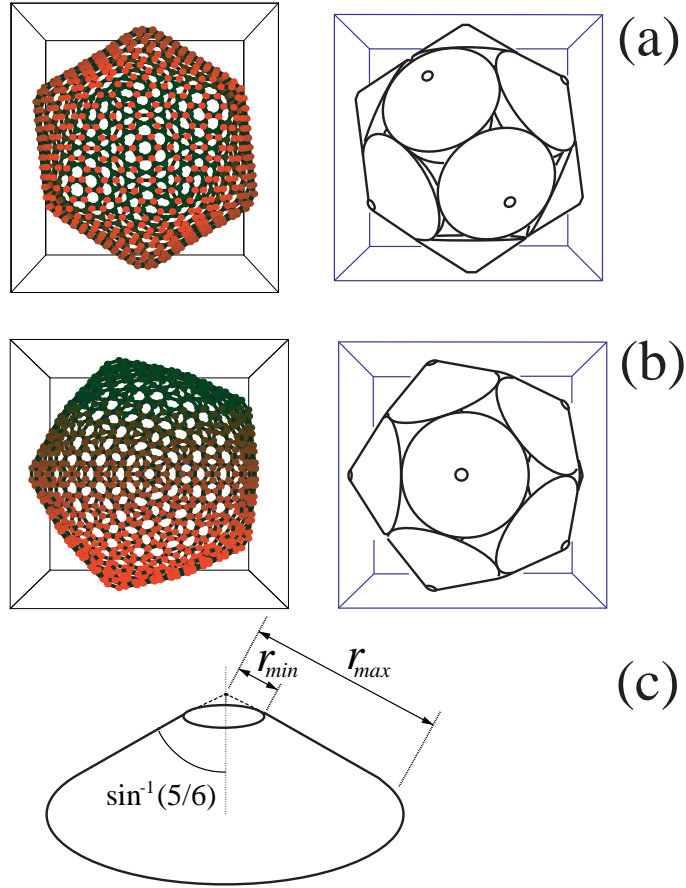,width=9cm}
}
\caption{Panel (a): Equilibrated, minimal energy $(m=4,n=0)$ shape (left) and its 
approximation by twelve conical frusta 
whose smaller bases centers are situated at the vertices of an icosahedron (right).
Panel (b): Same as in panel (a), only viewed from a different point. 
Panel (c): Elementary building block of the shapes shown in right 
columns of panels (a) and (b) - conical frustum whose half-angle is $\sin ^{-1} (5/6)$. The 
quantities relevant to Eq. (\ref{eq:conenerg}) are denoted.}
\label{fig:fig3}
\end{figure}

Icosahedral fullerenes are ($m,n=0$) subset of shapes considered in 
Sec. \ref{sec:sec1}. They are more complex than the infinite carbon 
nanotubes from the standpoint of elasticity 
theory since they contain twelve pentagonal disclinations situated 
at vertices of an icosahedron. The excess energy of icosahedral fullerene 
relative to a piece of infinite graphene sheet with the same number of atoms is 
thus a result of both bending (and possibly stretching) of 
the sheet {\em and} the energy 
required to create disclinations. In order to calculate the bending contribution 
to the energy, one should have some information on the geometry of the minimal 
energy shape of the icosahedral fullerene. The pentagonal disclination 
can be constructed in a graphene plane, starting 
from the center of a hexagon, and cutting the plane in two directions that make an 
angle of 60 degrees - these cuts can be easily seen in the geometrical 
constructions in Fig. \ref{fig:fig1}.
The smaller, cut-out piece of graphene is discarded and 
the remaining (larger) piece of graphene is then folded to 
rejoin the edges of the cut. The shape of minimal energy 
that the remaining piece of graphene adopts is a cone, since it 
costs less energy to bend the C-C bonds than to stretch them (it is convenient 
to imagine a circular piece of graphene plane with a hexagon in 
its center). The shape will 
in fact be a truncated cone (or a {\em conical frustum}), since the pentagonal 
ring at the top of the cone will be flat and parallel to the 
larger base of the cone. The precise description of the shape 
depends on the values of 
two-dimensional Young's modulus and bending rigidity. For 
materials like graphene (whose stretching requires much more energy 
than bending), the conical frustum is 
an excellent approximation of the minimal-energy shape. From 
purely geometrical considerations, the half angle of the cone is 
easily shown to be $\sin ^{-1}(5/6)$. Shapes of 
crystalline, hexagonally coordinated membranes with pentagonal 
disclinations have been studied in Ref. \cite{Seung}, and many 
of the features discussed there apply also to the case of graphene. 
The icosahedral fullerene is a bit more complicated from a single cone since 
it contains {\em twelve} pentagonal carbon rings. It seems plausible that the fullerene 
shape should be well described by {\em union} of twelve truncated 
cones. The radius of the larger base of a particular 
cone is restricted by the presence of other cones. To illustrate 
these geometrical considerations, in Fig. \ref{fig:fig3} 
the shape of $(m=4, n=0)$ icosahedral fullerene is displayed (left column of images), 
which is in a minimum of total energy, obtained numerically as discussed in 
Sec. \ref{sec:sec2}. The images in the right column of Fig. \ref{fig:fig3} 
display a union of twelve cones with half angles of 
$\sin ^{-1}(5/6)$ whose larger bases just touch. Note that this geometrical 
construction is an excellent approximation to the {\em calculated} 
minimal energy shape. Its small drawback is that its total area (without 
inclusion of the areas of the larger bases of the cones) is smaller 
from the total area of the fullerene shape which is manifested 
is Fig. \ref{fig:fig3} by the appearance of ''holes''. Note also that 
the regions where these holes appear are almost flat in the numerically 
calculated minimal energy shape of the fullerene, and their contribution 
to the bending energy should thus be small.

\begin{figure}[ht]
\centerline{
\epsfig {file=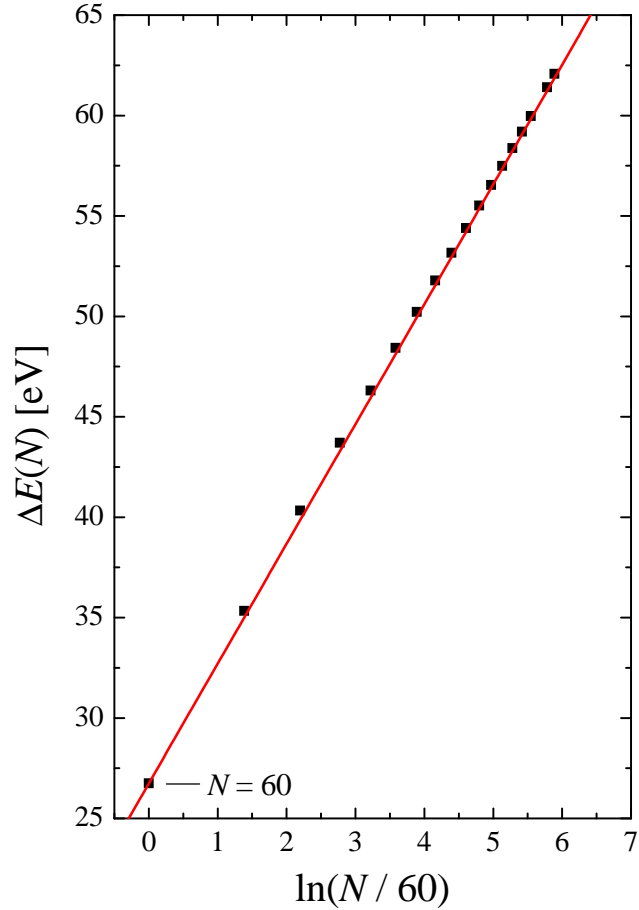,width=9cm}
}
\caption{Numerically calculated excess energies of icosahedral fullerenes (squares) 
and the prediction of Eq.(\ref{eq:delfulc60}) (line).}
\label{fig:fig4}
\end{figure}

Having such a good representation of the shape, it is possible to directly 
proceed to the calculation of its bending energy. In the continuum 
elastic theory of membranes \cite{LandauL}, the bending energy ($H_b$) is 
represented as
\begin{equation}
H_b = \frac{1}{2} \int_{\rm{Surface}} dS \left ( c_0 \kappa^2 + 2 c_G \gamma \right ),
\label{eq:elastica2}
\end{equation}
where $c_0$ is the bending rigidity discussed in the previous subsection, 
$c_G$ is the Gaussian rigidity, $dS$ is 
the infinitensimal element of the shape surface, and $\kappa$ and 
$\gamma$ are the mean and Gaussian curvature of the surface, respectively 
(if $R_1$ and $R_2$ are the principal radii of curvature, $\kappa = R_1^{-1} + R_2^{-1}$, 
and $\gamma = (R_1 R_2)^{-1}$). Note that for a cylindrical surface, 
$\gamma = 0$ and $\kappa = 1/R_{cyl}$ ($R_1=R_{cyl}$, $R_2 = \infty$) in 
every point of the surface, and the total bending energy 
integrates to $H_b / A = c_0 \kappa^2 / 2$, where $A$ is the total area of 
the cylinder (without the bases). This is in agreement with 
Eq. (\ref{eq:elastica1}). For conical 
surfaces, the gaussian curvatures are also zero everywhere, so 
the total bending energy of the approximated fullerene shape will depend only on $c_0$.
Integrating Eq. (\ref{eq:elastica2}) over the surface of a cone whose 
half-angle is $\beta$, one finds that the bending energy of a cone 
($H_b^c$) is 
\begin{equation}
H_b^c = \frac{\cos^2 \beta}{\sin \beta} \pi c_0 \ln \left ( r_{max} / r_{min} \right ),
\label{eq:conegeneral}
\end{equation}
which for the cones with $\beta = \sin^{-1} (5/6)$ yields \cite{tersselast} 
\begin{equation}
H_b^c = \frac{11 \pi}{30} c_0 \ln \left ( r_{max} / r_{min} \right ),
\label{eq:conenerg}
\end{equation}
where the meaning of $r_{min}$ and $r_{max}$ is illustrated in panel (c) of 
Fig. \ref{fig:fig3} - these are the distances of the upper and lower bases of 
the conical frustum from the cone apex measured along the cone face. The value 
of $r_{min}$ can be easily evaluated due to the fact that the smaller base of 
the frustum is a pentagon of carbon atoms. There is some ambiguity, however, 
since the pentagon is to be approximated by a circle. If the {\em circumradius} 
of the pentagon is identified with the radius of the smaller base of the cone, 
one obtains that 
$r_{min} = 3a \sqrt{50 + 10\sqrt{5}} / 25 \approx 1.021 a$, which is the 
upper bound for $r_{min}$. The identification of the radius of the smaller 
base of the cone with the {\em inradius} of the pentagon yields 
$r_{min} = 3a \sqrt{25 + 10\sqrt{5}} / 25 \approx 0.826 a$, which is 
the lower bound for $r_{min}$. In any case, 
$r_{min} = f a$, where $f$ is a numerical factor between 0.826 and 1.021. 
The value of $r_{max}$ is half of the shortest distance between the two 
apices of the neighboring cones measured along the cone faces.
From Figs. \ref{fig:fig1} and \ref{fig:fig3}, one finds that 
$r_{max}=3 (m-2) a / 2 + r_{min}$, and by using Eq. (\ref{eq:totnum}) one obtains 
that 
\begin{equation}
\frac{r_{max}}{r_{min}} = \frac{3}{2f} \sqrt{\frac{N}{60}} + \frac{f-1}{f}.
\end{equation} 
It is obviously quite convenient to choose $f=1$, closer to the upper bound for 
$r_{min}$, so that the total excess energy of the icosahedral fullerene 
can be written as 
\begin{equation}
\Delta E (N) = 12\left[\lambda_5 + \frac{11 \pi}{60} c_0 \ln \left( \frac{N}{60} \right) 
+ \frac{11 \pi}{30} c_0 \ln \left( \frac{3}{2} \right) \right] + E_{holes}. 
\label{eq:deltaefull}
\end{equation}
The multiplicative factor of 12 in the above equation is due to sum of the 
energies of twelve cones that make the shape surface, and $\lambda_5$ is 
the {\em core energy} of the pentagonal disclination which contains the 
local effects associated with the atomic structure of carbon pentagon and 
its immediate neighborhood. The 
contribution of the part of the fullerene 
surface that was not covered by the union of cones to the total energy 
was denoted by $E_{holes}$. Note that this quantity should represent a
quite small correction to the total energy (see Fig. \ref{fig:fig3}) 
and is not expected to 
depend strongly on $N$, since the area of the ''holes '' is proportional to $N$, while 
their mean curvature is approximately proportional to $1/\sqrt{N}$ 
(inverse radius of the fullerene, see Eq. (\ref{eq:elastica2})). Although one 
could easily invent a nice geometrical model to estimate this contribution, 
it is possible to proceed further by noting that Eq. (\ref{eq:deltaefull}) can 
also be written as
\begin{equation}
\Delta E (N) = \Delta E (C_{60}) + 
\left [\frac{11 \pi}{5} c_0 \ln \left( \frac{N}{60} \right) \right ],
\label{eq:delfulc60}
\end{equation}
where $\Delta E (C_{60})$ is the total excess energy of C$_{60}$ for which 
the calculation with the Brenner potential predicts $\Delta E (C_{60})=26.744$ eV. 
Figure \ref{fig:fig4} displays the results of numerical minimization of total 
energy of icosahedral fullerenes together with the prediction of 
Eq. (\ref{eq:delfulc60}) with $c_0=0.863$ eV taken from the analysis 
of infinite carbon cylinders. It seems more sensible to choose 
this value of $c_0$ rather than the one 
predicted by Eq. (\ref{eq:atomic1}) which is 
appropriate for small curvatures, 
since close to the smaller bases of the truncated cones, condition 
$a/R \ll 1$, where $R$ is the radius of curvature is not fulfilled.
Thus, a usage of a somewhat larger value of $c_0$ obtained from the 
numerical analysis of infinite carbon cylinder should approximately 
account for the contribution of terms like $(a/R)^p$, $p>2$ to 
the bending energy of the shapes studied. Note, however, that 
the (infinitensimal) analytical value of $c_0$ (0.83 eV) and 
the one obtained from the study of infinite carbon cylinders (0.863 eV) 
are quite similar, and one could also chose to work with the 
analytical value of $c_0$ with comparable success (see below).

It can be seen in Fig. \ref{fig:fig4} that the agreement between the calculated 
values and those predicted by Eq. (\ref{eq:delfulc60}) is excellent (note 
that there are {\em no fit parameters}). Very small 
deviations can be mostly attributed to the fact that the regions around the 
larger bases of the cone are constrained by the continuity of the fullerene surface, 
and deviation from the cone shape is thus expected. It is to a certain extent 
surprising that the predictions of Eq. (\ref{eq:delfulc60}) are so precise, 
especially concerning the fact that the excess energy of C$_{60}$ figures in 
it as a parameter. Buckminsterfullerene is certainly not expected to be 
reliably described by the elasticity theory, since there are no atoms on 
the cone faces, only C-C bonds. Yet, the elasticity theory results maintain 
their meaning down to fullerenes of very small sizes and small 
number of atoms. By examining the energies of 
individual atoms in the icosahedral fullerenes it is possible to proceed 
a bit further and to estimate the value of the core energy, $\lambda_5$. 
For example, in $(m=4, n=0)$ fullerene 
shown in Fig. \ref{fig:fig3}, the energies of atoms in pentagonal rings 
are -6.959 eV, while atoms situated in hexagonal carbon rings have 
energies of about -7.38 eV, depending on their position (see Eq. 
(\ref{eq:modelpotential})). The energies of atoms 
in pentagonal rings do not change significantly with increasing the 
size of the fullerene, which is a consequence of the {\em local} nature 
of this energy. All this suggests that 
$\lambda_5 = 5[-6.9587 - (-7.3949)]$ eV $= 2.18$ eV, where -7.3949 eV is 
the energy per atom in the infinite graphene plane, predicted by the 
Brenner potential. Equation (\ref{eq:delfulc60}) does not 
depend (formally) on this quantity, but it can be calculated from 
Eq. (\ref{eq:deltaefull}) which yields $\lambda_5^{elast} = 1.83$ eV 
($E_{holes}$ was set to zero), in a 
quite nice agreement with the number obtained from the detailed 
atomic description of the icosahedral fullerenes, especially 
in a view of all the details involved in the approximation of 
the fullerene shape. This also confirms that $E_{holes}$ is indeed 
negligible part of the total energy. 

\subsection{Capped carbon nanotubes, elongated icosahedral fullerenes}
\label{elongated}

Using the same reasoning outlined in the previous subsection, it is easy to construct 
an approximate shape of the elongated icosahedral fullerene ($n \ne 0$). 
These can be approximated by the union of twelve conical frusta {\em and} 
a cylinder whose length is $n a \sqrt{3}$, and radius $15ma / (2 \pi)$ (a 
finite-length piece of armchair (5$m$, 5$m$) carbon nanotube). 
The excess elastic energy can be thus written as
\begin{equation}
\Delta E (m,n) = 12 \left[ \lambda_5 + \frac{11 \pi c_0}{30}  
\ln \left(\frac{3 m}{2}\right) \right] + \frac{2 \pi^2 c_0}{5 \sqrt{3}} 
\frac{n}{m}  + E_{holes},
\label{eq:cappeden}
\end{equation}
which could also be parameterized by the total number of atoms, $N$, and 
$n$. The first term in the expression 
for energy in Eq. (\ref{eq:cappeden}) 
is the sum of the local and bending energies associated with the twelve 
equal pentagonal disclinations (conical frusta), and the second one corresponds 
to the bending energy of the finite length cylinder. The third term is 
again the energy associated with the ''holes'' in the geometrical construction, 
as discussed in the previous subsection. Note that for $n=0$, 
Eq. (\ref{eq:cappeden}) reduces to Eq. (\ref{eq:deltaefull}). Although 
the geometrical construction yielding Eq. (\ref{eq:cappeden}) is quite simple, 
it has a small drawback in that the areas of the cones and of the cylinder 
in between them overlap to a small extent. Angles between the surfaces 
of the cones and the cylinder at the points in which the entities nearly 
touch are small but nonvanishing, which is another 
drawback of the proposed construction. More elaborate constructions are 
possible, but are not needed (Fig. \ref{fig:fig6}). 
\begin{figure}[ht]
\centerline{
\epsfig {file=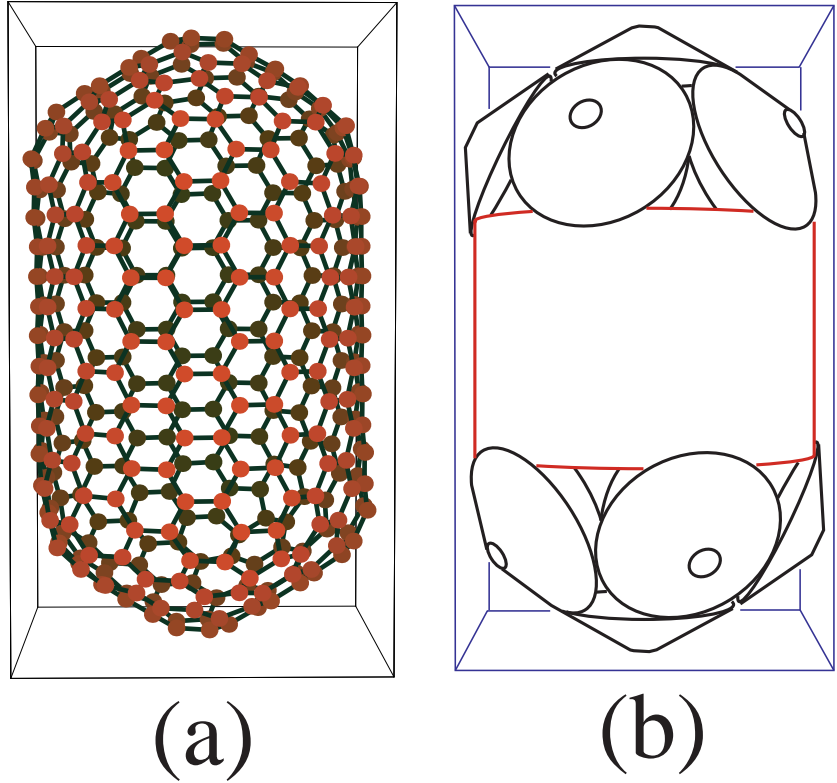,width=11cm}
}
\caption{Panel (a): Numerically calculated equilibrium shape of $(m=2, n=4)$ elongated 
icosahedral fullerene. Panel (b): Approximation of shape using geometrical construction 
leading to Eq. (\ref{eq:cappeden}).}
\label{fig:fig5}
\end{figure}
Equation \ref{eq:cappeden} can be approximated by 
\begin{equation}
\Delta E (m,n) = 
\Delta E (C_{60}) + 
\left [\frac{22 \pi}{5} c_0 \ln \left( m \right) 
+ \frac{2 \pi^2 c_0}{5 \sqrt{3}} 
\frac{n}{m} \right ], 
\label{eq:fullelong}
\end{equation}
which simply states that the excess energy of the shape is equal to the 
sum of the excess energy of the icosahedral fullerene and the 
bending energy of the cylinder inserted in between its two ''halves'' 
(see Fig. \ref{fig:fig5}). This equation is, however, an approximation, 
since it is based on the assumption that the excess energies associated 
with the holes in the geometrical constructions are the same for the 
icosahedral fullerene and the elongated icosahedral fullerene, which is 
not exactly the case (compare Figs. \ref{fig:fig5} and \ref{fig:fig3}). It furthermore 
contains an assumption that the elastic energies of holes do not 
depend on $m$ and $n$ integers characterizing the structure. Nevertheless, 
an excellent account of the numerical data is obtained which can be 
seen from Fig. \ref{fig:fig6}. Note again that the analytical results 
presented in Fig. \ref{fig:fig6} represent the predictions of the 
elasticity theory {\em without} fit parameters.
\begin{figure}[ht]
\centerline{
\epsfig {file=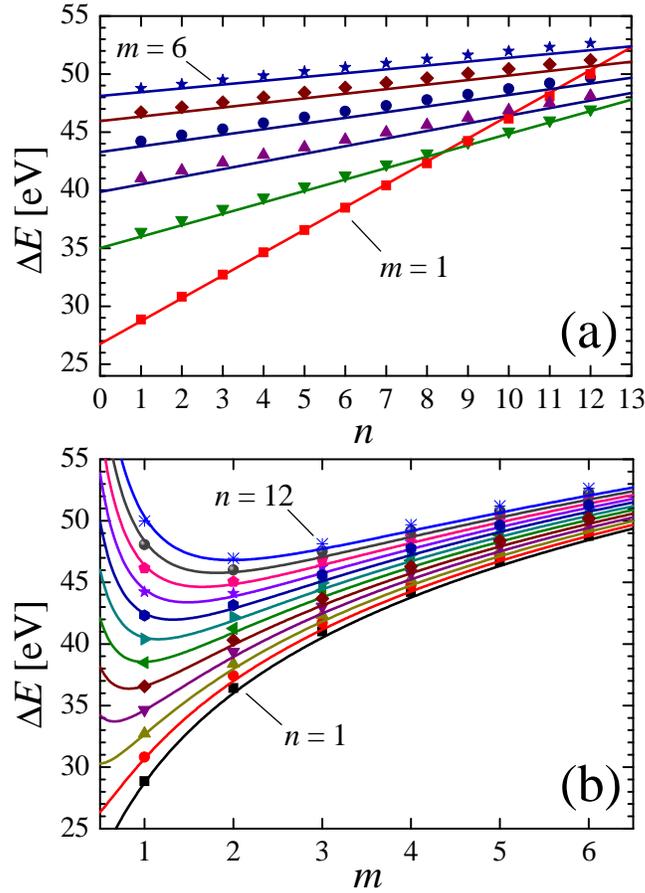,width=9cm}
}
\caption{Panel (a): Excess energies of elongated icosahedral fullerenes 
(capped carbon nanotubes) as a function of elongation number 
($n$) with $m$ as a parameter of the 
curves. Full symbols correspond to numerically calculated excess energies, 
while lines correspond to the predictions of elasticity 
theory in Eq. (\ref{eq:fullelong}). Panel (b): Same data as in 
panel (a) with the roles of $m$ and $n$ numbers interchanged.
}
\label{fig:fig6}
\end{figure}

\subsection{Contracted icosahedral fullerenes}
\label{squeezed}

\begin{figure}[ht]
\centerline{
\epsfig {file=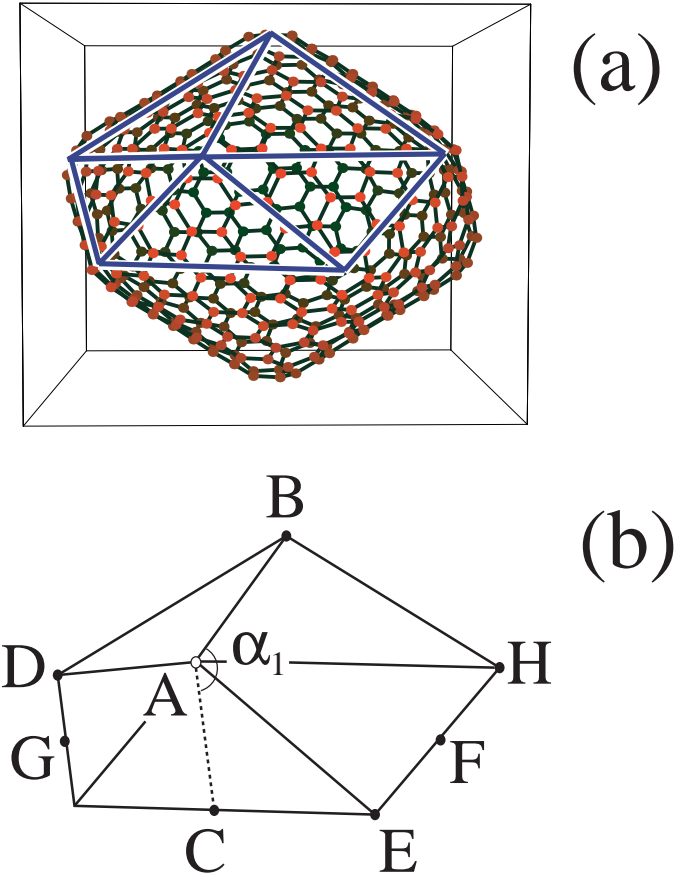,width=9cm}
}
\caption{Panel (a): Equilibrium shape of ($m=3$, $n=-2$) contracted 
icosahedral fullerene. The characteristic pyramid-like shape is 
denoted by full lines. Panel (b): The pyramid-like shape with 
denotations discussed in the text.}
\label{fig:fig7}
\end{figure}

The contracted icosahedral fullerenes are more complicated from the shapes 
considered thus far. Construction of their geometrical 
approximation is not simple mostly due to specific geometrical 
constraints that the ten pentagonal disclinations situated around the 
equator of the shape are subjected to.  When $n$ 
vanishes, the area around each of the disclinations is well 
represented by a cone with an angle of $2 \sin ^{-1} (5/6)$ which 
was clearly demonstrated in subsection 
\ref{icos}. When $n$ decreases (i.e. 
increases in absolute value), the two large pentagons 
parallel to the shape equator whose vertices coincide with five 
pentagonal disclinations (see Fig. \ref{fig:fig1}) approach each other. 
This constrains the shape of the area around these 
pentagonal disclinations. As the cone was such an useful 
approximation of the area around pentagonal disclinations 
for shapes studied in subsections \ref{icos} and 
\ref{elongated}, it seems reasonable to try to describe 
the contracted icosahedral fullerenes also as union of cones, 
ten (those arranged around the equator of the shape) of which 
belong to one category, and two (the ones at the 
two poles of the shape) to another. The two cones at the 
poles are easily constructed as in subsections \ref{icos} and 
\ref{elongated} - these have the angles of $2 \sin ^{-1} (5/6)$. 
The area around ten remaining disclinations is dominated 
by two opposing geometrical constraints. As $n$ decreases, the 
angle subtended by a line that connects a particular disclination 
(point A in panel (b) of Fig. \ref{fig:fig7}) 
and the one at the closest pole (point B) and a 
line that connects 
the disclination with the midpoint of a line drawn between 
the two nearest disclinations in the neigbouring 
pentagonal ring of disclinations (point C) also decreases 
- this angle is denoted with $\alpha_1$ in panel 
(b) of Fig. \ref{fig:fig7}. On the other hand, the angles 
that are subtended by the lines $\overline{\rm{DA}}$ and 
$\overline{\rm{AE}}$ or $\overline{\rm{FA}}$ and $\overline{\rm{AG}}$ 
(see Fig. \ref{fig:fig7}) {\em increase} when $n$ decreases. It is thus 
clear that the area around the disclination cannot be represented by 
a {\em single} cone. Nevertheless, it seems reasonable to approximate 
this area by a union of pieces of surfaces of two cones, each of them 
satisfying one of the two opposing geometrical constraints. 
The angle of one of the cones is about $\alpha_1$, which is 
given by
\begin{equation}
\alpha_1 (n) = \cos^{-1}\left[ \frac{1}{2 \sin(\pi/5)} \right ] + 
\cos^{-1} \left [ \frac{\sqrt{3}m \tan(\pi/10)}{3m - 2n} \right].
\end{equation}
The angle of the second cone ($\alpha_2$) that is searched for 
is larger than 
$\alpha_1$, and the contribution of this half-cone to the 
total elastic energy of the area surrounding the disclination 
is thus smaller (see Eq. (\ref{eq:conegeneral})). Nevertheless, the second 
cone is important since it contains a part of the total area 
surrounding the disclination, thus reducing the area available 
to the cone with angle $\alpha_1$. When $n=0$, both $\alpha_1$ 
and $\alpha_2$ angles should be $2 \sin^{-1} (5/6)$. 
Note that when $n$ decreases, 
angle $\alpha_2$ changes in a significantly smaller 
range than $\alpha_1$, since it is importantly 
constrained by the fixed angle between lines $\overline{\rm{DA}}$ and 
$\overline{\rm{AH}}$ (see Fig. \ref{fig:fig7}). Thus, I shall neglect the 
change of $\alpha_2$ with $n$ and set it to a constant
\begin{eqnarray}
\alpha_2 &=& 4 \sin^{-1} (5/6) - \alpha_1 (n=0) \nonumber \\
&=&  4 \sin^{-1} (5/6) - \cos^{-1}\left[ \frac{1}{\sin(\pi/5)} \right ] -  
\cos^{-1} \left [ \tan(\pi/10) \right] \nonumber \\
&\approx& 2.04 \sin^{-1}(5/6), 
\label{eq:alfa2}
\end{eqnarray}
which is obviously quite close to $2 \sin^{-1}(5/6)$, as it 
should be, and which was 
chosen so that $\alpha_2 + \alpha_1 (n=0) = 4 \sin^{-1} (5/6)$, since 
I am further going to assume that each of the two cones takes exactly a 
half of the total area available to the disclination. Together with the choice 
for $\alpha_2$ in Eq. (\ref{eq:alfa2}), this assumption shall 
yield the elastic disclination energy in the limit when $n=0$ equal 
to that characteristic of icosahedral fullerenes studied in subsection 
\ref{icos}. Note that the {\em axes} of the two cones do not need 
to be identical - the two pieces of the conical surfaces can be rotated 
with respect to each other at will. It is also irrelevant how the 
particular cone (half)surface is constructed as long as the pieces of it are bounded by 
the arcs of the two bases subtending the same angle from the cone axis 
and the two lines (along the cone face) that connect 
the arc ends in the upper and lower base. Thus, there is a significant 
freedom in constructing a shape around the disclination, but for all 
of the thus constructed shapes the buckling energies are equal. 
Note also that for $n \ne 0$ points 
D, B, H, and E do not lie in a plane. The proposed construction is 
admittedly rather approximate. For 
example, one could object that the fractions of the total area belonging 
to each of the two cones should also (at least slightly) change with 
$n$. Nevertheless, the construction accounts for 
{\em most important} geometrical constraints 
imposed on the area around pentagonal disclinations situated around the 
equator of the shape due to the presence of neighboring disclinations. 
The exact answers to all of the objections regarding 
the proposed construction that 
could be easily put forth would require 
a solution of the nonlinear problem in the theory of elasticity, which 
is not the aim of this study - the aim is to obtain a simple insight 
in the energetics of the shapes using the simplest possible 
application of the elasticity theory, and to obtain, preferably 
analytic, expressions for the excess energies of such shapes.

Following the reasoning presented in the 
previous paragraph and using Eq. (\ref{eq:conegeneral}) for the calculation 
of bending energy of a cone with arbitrary half-angle, the total excess 
energy of the proposed geometrical 
approximation to the contracted icosahedral shapes is 
\begin{equation}
\Delta E(m,n) = 12 \lambda_5 + 2 E_{pole}(m,n) + 10 E_{equator} (m,n) + E_{holes},
\label{eq:squeezenerg}
\end{equation}
where 
\begin{equation}
E_{pole} = \frac{11 c_0 \pi}{30} \ln ( 3m - q ),
\end{equation}
\begin{equation}
E_{equator} = \frac{c_0 \pi}{2} \ln (q) \left \{
\frac{\cos ^2 [\alpha_1 (n) /2]}{\sin [\alpha_1 (n)/2]}
+
\frac{\cos ^2 (\alpha_2 /2)}{\sin (\alpha_2/2)}
\right \},
\end{equation}
and
\begin{equation}
q = \frac{\sqrt{ 9m^2 + 3(3m-2n)^2 }}{4}.
\end{equation}
The last parameter ($q$) is simply a half of the distance (in units of $a$) 
between the two nearest neighboring disclinations situated at different pentagonal 
rings of disclinations around the equator of the shape (points A and E in 
Fig. \ref{fig:fig7}).

\begin{figure}[ht]
\centerline{
\epsfig {file=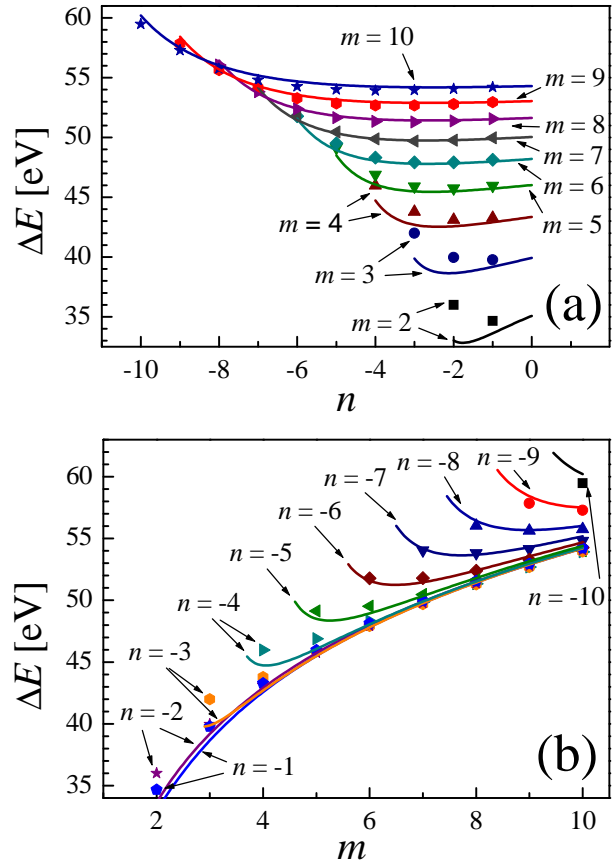,width=9cm}
}
\caption{Panel (a): Excess energies of contracted icosahedral fullerenes 
as a function of $n$ with $m$ as a parameter of the 
curves. Full symbols correspond to numerically calculated excess energies, 
while lines correspond to the predictions of Eq. (\ref{eq:squeezenerg}). 
Panel (b): Same data as in panel (a) with the roles of $m$ and $n$ 
numbers interchanged.
}
\label{fig:fig8}
\end{figure}

Numerically calculated excess energies of the shapes and the predictions of 
Eq. (\ref{eq:squeezenerg}) are shown in Fig. \ref{fig:fig8} and denoted by 
symbols and lines, respectively . In evaluating Eq. (\ref{eq:squeezenerg}), 
$E_{holes}$ was set to zero, $\lambda_5=1.83$ eV was taken from the 
analysis of icosahedral fullerenes in subsection \ref{icos}, and 
$c_0 = 0.863$ eV, as in the two previous subsections. Note again that 
there are no fitting parameters involved. 
The agreement of the predictions of the elasticity theory with the numerical 
results is striking. In spite of all the approximations involved in the 
geometrical construction of the equilibrium shape, Eq. (\ref{eq:squeezenerg}) 
both qualitatively and quantitatively accounts for the numerical data. 
The largest disagreement is found for the smallest shapes ($m=2$), but the 
curves correctly account for the appearance of minimum in excess energies for some 
finite $n$ (panel (a) of Fig. \ref{fig:fig8}), and for the increase in 
excess energies when $|n|$ becomes comparable to $m$. It is worth 
noting that although the number of atoms in $n=-m$ structures is smaller 
than in $(m,n=0)$ structures (regular icosahedral fullerenes) by 20 $mn$ 
(see Eq. (\ref{eq:totnum})), the total 
excess energies are still significantly larger in such structures, which 
is a consequence of an energetically unfavorable geometrical positions 
occupied by the ten disclinations around the equator of the shape. This 
consideration can also explain the minimum in excess energies obtained 
for some $n$ when $m$ is fixed - as the elongation number $n$ decreases, the number 
of atoms in the structure decrease, but the contribution to excess energy 
due to ten disclinations around the equator increase. The combined 
consequences of these two effects lead 
to the appearance of a minimum in excess energy as a function of $n$. 
Of the three types of shapes studied, the contracted icosahedral 
fullerenes typically have the largest excess energies per atom. For 
example, the excess energy of ($m=10$, $n=-10$) contracted fullerene 
($\Delta E=59.47$ eV) that contains 4000 atoms is significantly {\em larger} 
from the excess energy of ($m=9$, $n=0$) regular icosahedral fullerene 
($\Delta E=53.16$ eV) that contains {\em larger} number of atoms ($N=4860$).

\section{Summary and conclusion}
\label{sec:sec4}
It has been shown that the insight in the geometry of the equilibrium shapes 
can be extremely useful in estimating their (excess) energies. Such an 
approach does not require detailed description of the interatomic interactions 
characteristic of a shape or molecule in question, instead it relies 
exclusively on the knowledge of elastic parameters of the material that 
the shape (shell) is made of. In the studies presented in this 
article, the parameters that were necessary for reliable estimation of 
energy were the local energy of pentagonal ring of carbon atoms ($\lambda_5$), 
and the bending rigidity of graphene plane ($c_0$). Both of these 
parameters derive from the quantum-mechanical nature of bonding of 
atoms (see Eq. (\ref{eq:atomic1})) that make the shape. Surprisingly 
enough, it has been demonstrated that the application of the simplest 
theory of elasticity yields reliable results even for shapes 
made of a small number of atoms, i.e. in the regime where its applicability 
is not to be expected. For all of the shapes considered, the knowledge of 
the two-dimensional Young's modulus of graphene \cite{LandauL} was not needed, which 
suggests that in such shapes almost all of the elastic energy is 
of the {\em bending} type, while the energies associated with stretching are 
negligible \footnote{This, however, may change for extremely large fullerene 
molecules where the effects associated with the sharpening of the 
shape ridges may be of importance \cite{Lobkovsky,Witten}}. For shapes 
in which the geometrical constraints 
on the shape are strong (as in the case of contracted icosahedral fullerenes), 
the application of elasticity theory may be involved and complicated. 
Nevertheless, the identification of the constraints and their effects 
on the allowed shapes yields an additional insight in the energetics 
of the shapes and its behaviour within a certain class of shapes 
studied.

\end{document}